\documentclass{sig-alternate-05-2015}

\usepackage{graphicx}
\usepackage{amsmath}   
\usepackage{mathtools}  
\usepackage{amsfonts}  
\usepackage{url}     
\usepackage{color}  
\usepackage[usenames,dvipsnames]{xcolor} 
\usepackage{hyperref} 
\usepackage{threeparttable} 
\usepackage{amssymb} 
\usepackage{multicol} 
\usepackage{floatrow} 
\usepackage{booktabs} 
\usepackage{minipage-marginpar} 

\begin{document}
%
\conferenceinfo{Cyber and Information Security Research Conference}{2016 Oak Ridge, TN}
\CopyrightYear{2016} 
\crdata{978-1-4503-3345-0/15/04\\
}  

\title{
GraphPrints: Towards a Graph Analytic Method for Network Anomaly Detection
\titlenote{\tiny{This manuscript has been authored by UT-Battelle, LLC under Contract No. DE-AC05-00OR22725 with the U.S. Department of Energy.  The United States Government retains and the publisher, by accepting the article for publication, acknowledges that the United States Government retains a non-exclusive, paid-up, irrevocable, world-wide license to publish or reproduce the published form of this manuscript, or allow others to do so, for United States Government purposes.  The Department of Energy will provide public access to these results of federally sponsored research in accordance with the DOE Public Access Plan \url{http://energy.gov/downloads/doe-public-access-plan}. }}
}

\numberofauthors{1}
\author{
\end{tabular}
\begin{tabular}{c}	
	Christopher R. Harshaw,$^1$ 
	Robert A. Bridges,\\ 
	Michael D. Iannacone, 
	Joel W. Reed, 
	John R. Goodall\\
		\email{\href{crh7@rice.edu}{crh7@rice.edu}}, 
		\{\email{\href{mailto:brigesra@ornl.gov}{bridgesra}},	
		\email{\href{mailto:iannaconemd@ornl.gov}{iannaconemd}}, 
		\email{\href{mailto:reedjw@ornl.gov}{reedjw}}, 	
		\email{\href{mailto:jgoodall@ornl.gov}{jgoodall}}\}\{@ornl.gov\} 
\end{tabular}
\begin{tabular}{c}
		\affaddr{$^1$Oak Ridge Institute for Science \& Education}\\		    
		\affaddr{Oak Ridge, TN 37831}
		\affaddr{\&}\\
		\affaddr{Rice University}
		\affaddr{Houston, TX 77005}
\end{tabular}
\begin{tabular}{c}
	\affaddr{Computational Sciences and Engineering Division}\\
	\affaddr{Oak Ridge National Laboratory}\\
    \affaddr{Oak Ridge, TN 37831}
}

\date{\today}
\maketitle
\begin{abstract} 
This paper introduces a novel graph-analytic approach for detecting anomalies in network flow data called \textit{GraphPrints}. 
Building on foundational network-mining techniques, our method represents time slices of traffic as a graph, then counts graphlets\textemdash small induced subgraphs that describe local topology. 
By performing outlier detection on the sequence of graphlet counts, anomalous intervals of traffic are identified, and furthermore, individual IPs experiencing abnormal behavior are singled-out. 
Initial testing of GraphPrints is performed on real network data with an implanted anomaly. 
Evaluation shows false positive rates bounded by 2.84\% at the time-interval level, and 0.05\% at the IP-level with 100\% true positive rates at both. 
\end{abstract} 
\section{Introduction}
\label{intro}
As government, industry, and private enterprises are increasingly dependent on information technology, adequate cyber defensive capabilities are of the utmost importance. 
Currently, defending networked computing assets relies on two primary mechanisms \textemdash automated signature-based detection systems and manual analysis of cyber data. 
While essential, these alone are insufficient.  
Signature-based detection methods (such as anti-virus software, intrusion detection systems, and firewalls) identify malicious traffic using heuristics that are usually handcrafted; hence, such methods are only effective when encountering a previously analyzed attack. 
In order to gain traction in understanding network activity, security analysts turn to semi-manual investigation of network data (e.g., firewall logs, packet capture data, network flows, system logs), but are met with an abundance of data and a scarcity of information; 
for example, flow data alone from a small network will constitute 10-100 million records per day. 
Anomaly detection methods, which seek to single out unexpected events, propose a much needed compliment to current methods, as they pinpoint noteworthy events for further investigation and hold the promise of detecting never-before-seen attacks. 

In this paper, we introduce \textit{GraphPrints}, a novel network anomaly detection algorithm that uses graph analytics to identify qualitative changes in network traffic at the IP and whole network level. 
We concentrate on network flows\textemdash records of communication between a source and destination IP address. 
More specifically, flows give the metadata (timestamp, IPs, ports, protocol, etc.) describing communication between a pair of IPs over a small time window (see inset table in Section~\ref{workflow}). 
As such, network flows are a primary data source for monitoring, diagnosing, and investigating network traffic.
Our method divides network traffic into time slices that are naturally represented as a graph. 
More specifically, a graph $G = (V, E)$ is defined as a set of vertices, $V$ (representing entities, IPs in our case), and edges, $E$ (representing relationships between pairs of vertices\textemdash 
communication between IPs in our case.) 
The GraphPrints method mines \textit{graphlets}\textemdash small, induced subgraphs, which can be thought of as the building blocks of the graph that describe the local topology\textemdash then performs outlier detection to find those time windows of traffic that exhibit uncharacteristic graphlet counts.
In order to isolate the specific IPs engaging in unexpected traffic, for each vertex $v$ of a graph, GraphPrints also tracks the graphlet automorphism orbits containing $v$ (see Section~\ref{graphlets}), which characterizes the role $v$'s IP plays in communication patterns. 
This simultaneous detection capability not only identifies time windows of anomalous traffic, but also focuses the operator on those IPs in the time window exhibiting uncharacteristic behavior.  

Our hypothesis is that an unusual occurrence in network traffic will produce 
a detectable change in graphlet counts.  
To test this, we implement the multi-level detector on real network traffic with an implanted anomaly\textemdash a host engaging in bit torrent traffic
\textemdash and report promising results highlighting GraphPrints' potential (Section~\ref{eval}).

The GraphPrints method builds on foundational work of Milo et al.~\cite{milo2002network} and Pr{\v{z}}ulj et al.~\cite{prvzulj2004modeling} (among others) where graphlets and automorphism orbits are used for network classification, alignment, and comparison. 
To the best of our knowledge, this is the first work to propose a network anomaly detection algorithm for the dynamic graph setting based on graphlets and automorphism orbits. 
We note that the idea for multi-level anomaly detection on time-varying graph data follows contributions of Bridges et al.~\cite{bridges2015multi}, which create a generative model for detection on synthetic graph data at node, community, and whole graph levels. 
Additionally, a position paper of Halappanavar et al.~\cite{halappanavar2013towards} outlines a network-of-network design for cyber security applications and calls for graphlet analysis of flow data, but to our knowledge no implementation of these ideas has yet been pursued.


\section{GraphPrints Workflow}
\label{workflow}
In order to perform streaming anomaly detection based on graph analytics, network flow data must be observed, converted into a sequence of graphs, and fed to an algorithm for identifying aberrations in the graph data. 
This section outlines the steps used in our detection workflow from tapping network communication, to identifying unexpected time windows of network activity. 

To tap network flow data, we used ARGUS\footnote{\url{http://qosient.com/}} (the Audit 
\begin{minipage}{0.32\linewidth}
\noindent Record Generation and Utilization System),  an open source, real-time, network flow monitor.
The table to the rights contains an example of a single ARGUS flow 
\end{minipage}
\begin{minipage}{.19\linewidth}
\phantom{blah} 
\end{minipage}
\begin{minipage}{.48\linewidth}
	\phantom{aaa\\}
	\begin{tabular}{cc}
	\toprule	
	\textbf{Time} & 09:58:32.912\\
	\textbf{Protocol} & tcp \\
	\textbf{SrcIP} & 192.168.1.100 \\
	\textbf{SrcPort} & 59860 \\
	\textbf{DstIP} & 173.16.100.10 \\
	\textbf{DstPort} & 80 \\
	\textbf{TotBytes} & 1695088 \\
	\bottomrule
	\label{tab:flow}
	\end{tabular}
\end{minipage}\\
 record. Only the fields used in this paper are displayed. 


\subsection{Representation of Flow Data as Graphs}
\label{graph-creation} 
We seek to represent network flow data in a given time interval as a directed graph (digraph) $G$ using a vertex for each IP and letting flow records generate edges. 
To do this, we create a preliminary graph $H$ with vertices for IP addresses and, naturally, an edge for each flow. 
More specifically, upon observation of a flow record, a directed edge from the source IP's vertex to the destination IP's vertex is created, and we weight each edge by the flow's \textit{total bytes} record. 
Furthermore, we color an edge \textcolor{BlueViolet}{blue} if at least one of the ports is well known (i.e., less than 1024) and  \textcolor{BrickRed}{red} if both ports are not well known (1024 or greater).\footnote{The most common TCP and UDP traffic will have a destination (or source) port of a well-known port number, with the source (or destination) port being a randomly chosen ``high'' port number.  Traffic that has both ports outside of this range is typically either a less-common protocol, or a common protocol configured in an uncommon way.} 
Since multiple flows can occur between the same pair of IPs in the time interval, $H$ is a multi-digraph, meaning it admits multiple edges between any pair of vertices, and we aggregate the multi-edges to create the desired graph, $G$.
Specifically, for each source vertex $i$ and destination vertex $j$ of $H$, let  $B_{i,j}$ ($R_{i,j}$) be the sum of edge weights, total bytes, over the blue (red) edges from $i$ to $j$. 
If $B_{i,j} > R_{i,j}$ ($R_{i,j} \geq B_{i,j}$), create a blue (red) edge from $i$ to $j$ in $G$. 
Thus our final graph, $G$, has  colored, directed, and unweighted edges representing the collection of flows observed in the time interval. 

\subsection{Graphlet \& Orbit Vectors}
\label{graphlets}

\begin{figure}
\epsfig{file=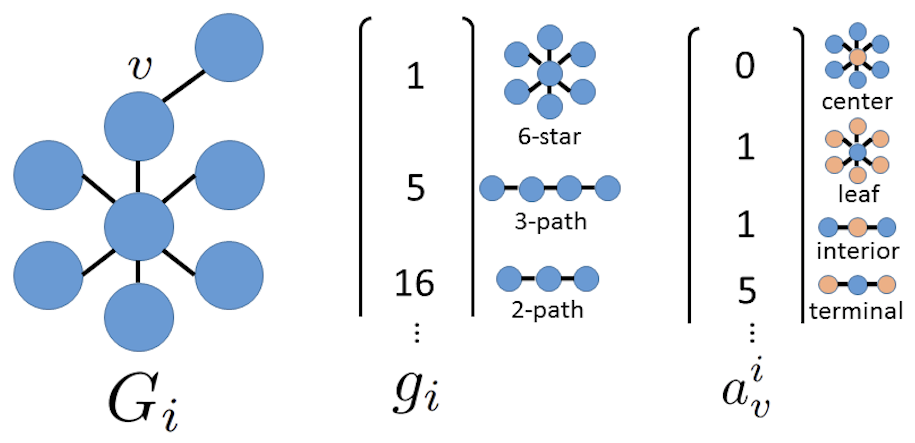, height=1.6in, width=3.35in}
\caption{Graphlet \& Orbit Counting Example}
\vskip -10pt
\label{fig:graphlet}
\end{figure}

To capture a network's behavior over time, we examine \textit{graphlets}\textemdash small, connected, induced subgraphs of the network.
Graphlets are an important network characteristic because they capture the local topology of the network. 
For instance, observing many stars or many paths (see Figure~\ref{fig:graphlet}) provides an understanding of the overall structure of the network~\cite{milo2002network, prvzulj2004modeling}. 
Given a graph $G_i$ from time window $i$, we count the occurrences of all graphlets up to size $k$, and store these graphlet counts in a \textit{graphlet degree vector}, denoted $g_i$. 
See Figure~\ref{fig:graphlet} for an example graph $G_i$ with graphlet degree vector $g_i$.
The notion of graphlets is easily extended to include colored vertices and edges as well as directed edges, as is our current interest. 
In this case, topologically equivalent graphlets with different colorings are considered different.   
While coloring allows encoding of contextual information, the number of graphlets grows combinatorially with the number of colors, which can pose issues of computational cost or memory usage. 

To capture a vertex's behavior over time, we will examine graphlet \textit{automorphism orbits}.\footnote{ 
The \textit{automorphism orbits} of $G=(V,E)$ form a partition of $V$ where vertices $u$ and $v$ are in the same orbit iff there exists a graph automorphism $f:V \rightarrow V$ that takes $u$ to $v$.} 
The automorphism orbits of a graph are sets of vertices that are symmetric in the graph. 
Intuitively, the automorphism orbits are the different roles a vertex can fill in that graph;  
for instance, a star has two automorphism orbits\textemdash one for the center vertex and one for the leaf vertices. 
The \textit{orbit vector of $v$} is a vector whose entries are the counts of the graphlet automorphism orbits in which  $v$ participates. 
Given a graph $G_i$ and a vertex $v$, we denote the orbit vector of $v$ in $G_i$ as $a_v^i$ (see Figure~\ref{fig:graphlet}). 
Orbit vectors are useful characteristics because they describe a vertex's extended neighborhood in a graph. 
For instance, a vertex who appears in the center of many stars is likely different from a vertex who appears always as a leaf.

To summarize the workflow up to this point, network flow data is observed in near real time, and is represented as a sequence of graphs, $G_1, G_2, \dots$. 
For a given graph $G_i$, graphlet vectors $g_i$ characterize the network at the graph-level (i.e., changes over the whole time-window of network activity) and the orbit vectors $a_v^i$ characterize the network at the vertex level. 
As an analogy, graphlet vectors are like handprints and orbit vectors are like fingerprints, hence the term \textit{GraphPrints}. 
There are several available tools to count graphlets. 
In this study, we used FanMod\footnote{\url{http://theinf1.informatik.uni-jena.de/motifs/}}~\cite{wernicke2006fanmod}. 
We note that in the FanMod implementation there is no extra computation to count graphlet automorphism orbits as this is already a step in the process of graphlet counting.

\subsection{Anomaly Detection Framework}
\label{ad} 
%
In order to perform detection at the graph level, we fit a multivariate Gaussian to the set of previously observed graphlet degree vectors, $\{g_i\}_{i = 1}^{n}$ using a technique that is robust to outliers. 
Because our historical network data may contain (discovered or undiscovered) anomalies, we use the \textit{Minimum Covariance Determinant} (MCD) method, which can identify the ``best fit'' mean and covariance in the presence of up to 50\% outliers. 
More specifically, the user provides a number $h$ between $n/2$ and $n$ of pure (i.e., non-anomalous) data points,  and the algorithm finds the ellipsoid of least volume that covers $h$ points.\footnote{Once the ellipsoid, $E = \{x: (x-\mu)^tQ(x-\mu)\leq 1\}$ ($\mu \in \mathbb{R}^p, Q \in \mathbb{R}^{p \times p}$ positive definite),  is found, the corresponding Gaussian has mean $\mu$, and covariance $\Sigma = Q^{-1}$.} 
This is equivalent to fitting a Gaussian distribution to $h$ of the data points, and omitting $n-h$ outliers; hence our distribution will not be skewed by up to $(n-h)/n$ percent of arbitrarily bad data points. 
Upon receipt of the next time-window's graphlet vector, $g_{n+1}$ we compute the Mahalanobis distance,\footnote{Given mean $\mu$ and covariance $\Sigma$, the Mahalanobis distance is defined of $x$ is $(x-\mu)^t\Sigma^{-1}(x-\mu)$ and is (inversely monotonically) equivalent to finding the vector's $p$-value.} 
which scores how anomalous the new vector is\textemdash high scores for very anomalous data, low scores for relatively common data, according to the Gaussian distribution. 
We reference interested readers to algorithmic details of MCD by the creators, Rousseeuw et al.~\cite{rousseeuw1999fast}. 
Our implementation used the SciKit-Learn MinCovDet module~\cite{scikit-learn}.\footnote{\url{http://scikit-learn.org/stable/modules/generated/sklearn.covariance.MinCovDet.html}}

To demonstrate the usefulness of orbit vectors in IP flow analysis, we characterize normal vectors via an unsupervised clustering algorithm ($k$-means), and use a vector's distance to cluster centers as an anomaly score.  
The $k$-means algorithm is an unsupervised learning algorithm that, given positive integer $k$, partitions the data into $k$ clusters via a greedy method. 
To build a detector, we first learn the cluster centers (centroids) by running $k$-means on a set of observed data points. 
Next, given a (newly observed) orbit vector, we determine its anomaly score by considering its distance to the nearest centroid. 
In this study, we used the gap statistic to choose the value of $k$ used in $k$-means~\cite{tibshirani2001estimating}.    
Details of our detection experiment and results are given in Section~\ref{eval}.

\section{Experimental Setup \& Results} 
\label{eval}
For a demonstration dataset, network flow data was collected from the main network switch of a small office building in a campus environment, during a typical workday.  
This traffic includes both wired and wireless subnets, as well as a small datacenter with many virtual machines (VMs), on a separate NAT-ed network. 
This collection of flows serves as ambient traffic. 
To create a known anomaly,  the ambient traffic is combined with a separately-recorded session of bit torrent flows from a single host. 
 As bit torrent participation is disallowed on the office network, the bit torrent traffic should be abnormal. 
Furthermore, we expect bit torrent traffic to demonstrate a similar topology to other, more concerning types of traffic, such as peer-to-peer botnet communication or distributed denial of service attacks, which would be problematic to generate in a real network setting.   
To implant the bit torrent traffic the flows' timestamps were offset, and the host and router IP addresses mapped to their analogues in the building's subnet. 
Finally, the bit torrent flows are shuffled into the ambient traffic respecting the time sequence of all the data. 
In total, this dataset included 10,507 IP addresses of which 2,795 IP addresses are within the building's subnets. 
These included 151 IPs on the ethernet network, 491 IPs on the wireless network, and 2,153 IPs in the datacenter. 

As discussed in Section~\ref{graph-creation}, we now represent the data as a sequence of graphs, $(G_i)$, with time windows of 31 seconds with one second overlap. 
In total, we observed 350 graphs, averaging 1,265 nodes and 4,901 edges per graph of which 76\% were red indicating more data was sent between high port connections than otherwise. 
Additionally, there were on average 4,929 non-empty flows per time interval; hence, most colored edges represent a single flow. 
Zooming in on the bit torrent traffic, we find 40 intervals (no. 278-317 in Figure \ref{fig:graph-lvl}) in which at least a single flow was attributable to bit torrent traffic.
Of these, the first 24 time windows contained roughly 15\% bit torrent flows, while the latter 16 intervals had only contained 2\% bit torrent flows. 
For testing, we consider these 24 intervals true positives at the graph level, and the vertex engaging in the bit torrent activity during those intervals a true positive at the node level. 
As the goal of an anomaly detector is to pinpoint abnormal events, we consider a substantial change in network activity to be a true positive, even if not necessarily malicious. 
Additionally, we are unaware of what, possibly bizarre, activity is present in the rest of the data.

Graphlets of size three and corresponding automorphism orbits are counted to create the sequence of graphlet degree vectors $g_i$ and a sequence of orbit count vectors for each vertex $v$, $a_v^i$ (see Section~\ref{graphlets}). 
Using the MCD algorithm (Section~\ref{ad}), we fit the initial Gaussian to the first 150 graphlet vectors. 
For each subsequent vector, we score anomalousness using Mahalanobis distance and re-fit the Gaussian to include the new data point. 
In this study, $h = 0.85 n$.
See Figure~\ref{fig:graph-lvl}, which gives a plot of the anomaly score for each time window. 
A suggested threshold is displayed in red, and is chosen to maximize the known true-positives while minimizing the number of other points over the threshold. 
Since the known anomalies are easily discriminated from the majority of the traffic, such a threshold obtains perfect true positive rate, zero false negatives, and we see 5 of 176 unknown anomalies detected. 
This bounds our graph-level false positive rate by 2.84\%.  
Although we have not proven causality, initial investigation of these unknown anomalies revealed an IP scan of a VM subnet; hence, these may indeed be true (but previously unknown) positives. 
Further investigation of these detected anomalies will be included in future work. 


For node-level detection, we randomly sample 40 IPs with probability proportional to their vertex's occurrence in the data, then cluster their orbit vectors' from the first 150 time intervals. 
As described in section~\ref{ad}, we use the gap statistic and choose $k=5$ for K-means clustering.
For the remaining 200 time intervals the distance of these 40 IPs' orbit vectors to the nearest cluster centroid is plotted on the top line of Figure~\ref{fig:node-lvl} (blue dots), while just below, the red dots represent the nearest-centroid distances of orbit vectors corresponding to the known anomalous vertex. 
As in the graph-level analysis, an appropriate threshold that detects all known anomalies is indicated by the green bars. 
We note that only 4 of 8000 vectors from the unknown time intervals are detected as anomalous, bounding our false positive rate by 0.05\%. 
Furthermore, the clear disparity in scores exhibited by the known anomalies permits zero false negatives at the node level also.

These preliminary results show that graphlet vectors can indicate \textit{when} anomalies occur in the network, and orbit vectors can identify \textit{where} these anomalies occur.

\begin{figure}
\epsfig{file=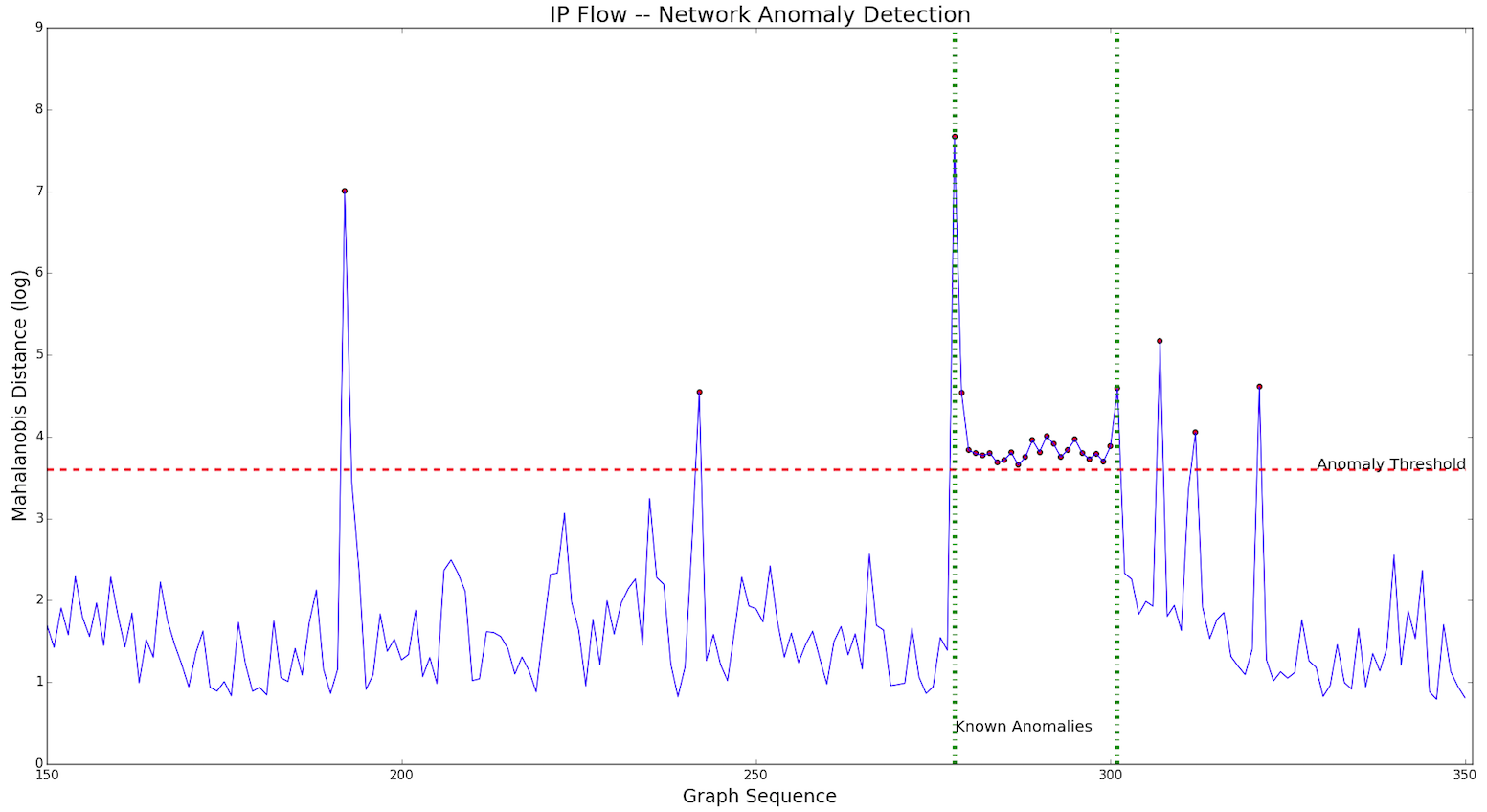, height=1.864in, width=3.35in}
\caption{Graph-Level Detection Results}
\label{fig:graph-lvl}
\floatfoot{\scriptsize{\textbf{Note:} Log-Mahalanobis distance of graphlet vectors shown.
Threshold chosen to maximize known anomalies, while minimizing detection of other points. True-positive rate = 100\%. False positive rate = 2.84\%.}}
\vskip -9pt
\end{figure}

\section{Conclusion and Future Work}
\label{conclusion}
Our evaluation of GraphPrints exhibited 2.84\% and 0.05\% false positive rates at the graph and node levels, respectively, with 100\% true positive rate at both. 
This confirms that noteworthy changes in network traffic are identifiable using only counts of 3-graphlets / orbits.   
Furthermore, the discovery of unexpected anomalies illuminates an important area of future research for GraphPrints\textemdash how to trace a detected event to (1) particular graphlet / orbit counts, and (2) the specific network traffic that is abnormal. 
As the GraphPrints method admits colored nodes, important contextual information\textemdash such as an IP's membership in a known subnet, ASN, or country code\textemdash  can be encoded via node coloring, and we expect future inclusion of this information to yield more informative results. 
Finally, we believe further investigation of the node-level clusters will allow us to characterize different types of user behavior; for example, perhaps one cluster is comprised of the orbit vectors associated with using email, while another those of web browsing. 
In summary, this work presents a graph analytic method with promising initial results for analyzing, detecting, and characterizing  network flow data, 
where noteworthy changes in network behavior are identifiable at multiple levels with exceptionally low false positive rates. 

\begin{figure}[t]
\epsfig{file=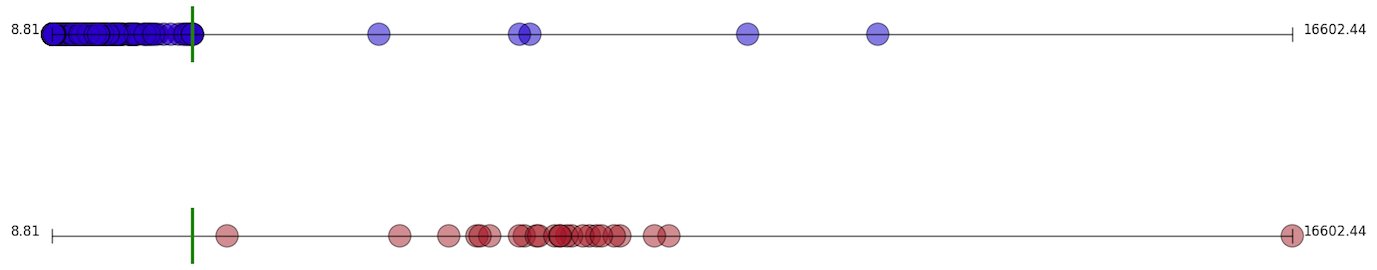, height=.6582in, width=3.35in}
\caption{Node-Level Detection Results}
\label{fig:node-lvl}
\floatfoot{\scriptsize{\textbf{Note:} Distance of orbit vectors 
 from nearest centroid depicted. Unknown vectors in blue dots. Known anomalies in red dots. Threshold chosen to maximize known anomalies, while minimizing other detected point. True-positive rate = 100\%. False positive rate = 0.05\%.}}
\vskip -9pt
\end{figure}

\section{Acknowledgments}
This material is based on research sponsored by: the U.S. Department of Homeland Security (DHS) under Grant Award Number 2009-ST-061-CI0001, DHS Science and Technology Directorate, Cyber Security Division (DHS S\&T/CSD) via BAA 11-02; the Department of National Defence of Canada, Defence Research and Development Canada (DRDC); the Kingdom of the Netherlands; and the Department of Energy (DOE). The views and conclusions contained herein are those of the authors and should not be interpreted as necessarily representing the official policies or endorsements, either expressed or implied, of: the DHS; the DOE; the U.S. Government; the Department of National Defence of Canada DRDC; or the Kingdom of the Netherlands.

\bibliographystyle{abbrv}
\bibliography{main} 
\end{document}